\documentclass{osa-article}

\journal{osajournal}

\articletype{Research Article}

\usepackage{lineno}

\begin{document}

\title{Propagation of temporal mode multiplexed optical fields in fibers: influence of dispersion}

\author{Wen Zhao,\authormark{1} Nan Huo,\authormark{1} Liang Cui,\authormark{1} Xiaoying Li,\authormark{1,*} Z. Y. Ou\authormark{2,+}}

\address{\authormark{1}College of Precision Instrument and Opto-Electronics Engineering, Key Laboratory of Opto-Electronics Information Technology, Ministry of Education, Tianjin University, Tianjin 300072, People’s Republic of China\\
\authormark{2}Department of Physics, City University of Hong Kong, Kowloon, Hong Kong, P. R. China}

\email{\authormark{*}xiaoyingli@tju.edu.cn}
\email{\authormark{+}jeffou@cityu.edu.hk} 



\begin{abstract}
Exploiting two interfering fields which are initially in the same temporal mode but with the spectra altered by propagating through different fibers, we characterize how the spectra of temporal modes changes with the fiber induced dispersion by measuring the fourth-order interference when the order number and bandwidth of temporal modes are varied. The experiment is done by launching a pulsed field in different temporal modes into an unbalanced Mach-Zehnder interferometer, in which the fiber lengths in two arms are different. The results show that the mode mismatch of two interfering fields, reflected by the visibility and pattern of interference, is not only dependent upon the amount of unbalanced dispersion but also related to the order number of temporal mode. In particular, the two interfering fields may become orthogonal under a modest amount of unbalanced dispersion when the mode number of the fields is $k\geq2$. Moreover, we discuss how to recover the spectrally distorted temporal mode by measuring and compensating the transmission induced dispersion. Our investigation paves the way for further investigating the distribution of temporally multiplexed quantum states in fiber network.
\end{abstract}

\section{Introduction}\label{Introduction}
The capacity and performance of communication system can be increased and improved by encoding information on the multiple degree of freedom of optical fields. For example, the multiplexing technique of wavelength, polarization and spatial modes greatly increases the information capacity of fiber optical communication system~\cite{Richardson2010Science}. For quantum information science (QIS), multiplexing is also an essential technology, which makes the quantum systems more scalable and practical due to increased channel capacity and tolerance to noise~\cite{Brecht-PRX2015,Erhard-Light2018,Fabre-Rev.Mod.Phys.2020}. While the mode basis in wavelength, polarization and spatial degree of freedom have been widely exploited in the past~\cite{Richardson2010Science,Erhard-Light2018,Fabre-Rev.Mod.Phys.2020}, the temporal modes (TMs) which provide a convenient orthogonal basis for studying light pulses of any temporal and spectral shape have not attracted much attention until recently~\cite{WalmsleyReview2020,Brecht-PRX2015}.

\begin{figure*}[h]
\centering
\includegraphics[width=\linewidth]{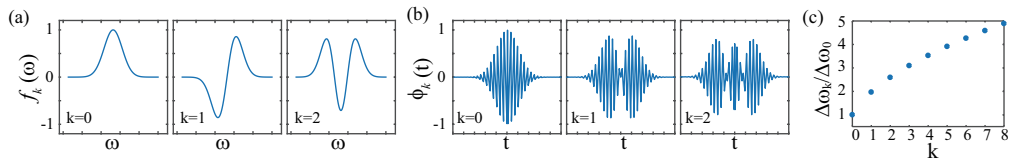}
\caption{First three Hermite-Gaussian modes in (a) the frequency domain $f_k(\omega)$ and (b) the time domain $\phi_k(t)$. (c) Frequency modes occupied by a temporal mode $\Delta\omega_k$ (normalized to $\Delta\omega_0$) varies with the order number k.}
\label{modes}
\end{figure*}

An example of such an orthogonal temporal mode (TM) basis is a set of Hermite-Gaussian modes. The first three Hermite-Gaussian modes of a TM basis are illustrated in Fig. \ref{modes}(a) and \ref{modes}(b). Because of the relation between time and frequency, frequency modes occupied by a TM increase with mode order number $k$ (see Fig. \ref{modes}(c)). So the TM multiplexing, at the cost of wavelength division multiplexing, can not increase channel capacity. This is similar to the spatial mode multiplexing in free space~\cite{Ningbo-NP2015}. However, TM analysis does offer a much more straightforward way with intrinsically decoupled modes in describing pulse-pumped parametric processes~\cite{yuen1989multimode,Brecht-PRX2015}, by which a variety of quantum states of light have been experimentally generated and used in quantum information processing, including quantum communication, quantum simulation, and quantum metrology.

It was recognized recently that the TMs of electromagnetic fields form a new framework for QIS~\cite{Brecht-PRX2015}, since they intrinsically span a high dimensional Hilbert space and lend themselves to integration into existing fiber communication networks. So far, a lot of efforts have been made to develop the tools for implementing photonic quantum information in TMs framework, such as the generation of quantum states by pulsed pumped parametric process for carrying information, and the manipulation of TMs to fulfill the function of multiplexing and demultiplexing. However, distribution of the temporally multiplexed quantum states in long distance optical fiber, which is a key for the success of quantum information processing in TMs framework, has not been done yet.
It is generally accepted that for a set of orthogonal TMs basis, the walk-off effect can be neglected because the TMs are centering at the same frequency and the group velocities for different orders of TM are equal. However, dispersion of transmission medium will inevitably affect the shape of pulsed field. As a result, the devices designed for accomplishing the functions of TMs demultiplexing and detection will be affected.
In this paper, as a first step for long-distance distribution of quantum states in higher-order TM, we study how the dispersion of optical fiber influence the TMs of electromagnetic fields by using an unbalanced Mach-Zehnder interferometer (MZI) and measuring the visibility of fourth-order interference.

The rest of the paper is organized as follows. In Sec. \ref{Theory}, we introduce the theory of how to characterize the spectral change of TMs with Hong-Ou-Mandel type fourth-order interference~\cite{HOM1987,Ou1999}. After briefly reviewing the fourth-order interference between two pulsed interfering fields, we analyze the fourth-order interference of an unbalanced MZI when its input is in a single TM but with different orders. In Sec. \ref{Experiment and Results}, we describe our experiments carried out by using the unbalanced MZI, in which the fiber lengths of two arms are different. We show how the transmission fiber induced dispersion affects the spectral profiles of the TM with different order numbers by measuring the pattern and visibility of fourth-order interference. In Sec. \ref{Discussion}, we discuss how to measure the dispersion of transmission medium by using the fourth-order interference of  unbalanced MZI, which can be used to cancel the dispersion induced spectral change of TM. Finally, we conclude in Sec. \ref{Conclusion}.

\section{Theory}\label{Theory}
It is well-established that the visibility of the Hong-Ou-Mandel (HOM) type fourth-order interference depends on the degree of mode matching between two interfering fields ~\cite{HOM1987,Ou1999,Ma2015}. Using this property of HOM interference, we can characterize how the spectra of temporal modes (TMs) change with the dispersion induced by transmission media.

\begin{figure*}[h]
\centering
\includegraphics[width=\linewidth]{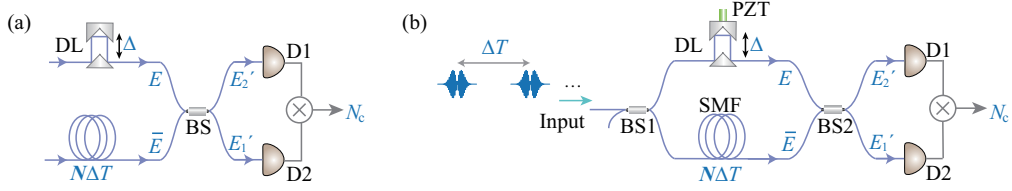}
\caption{(a) Schematic diagram of a Hong-Ou-Mandel (HOM) interferometer. (b) Conceptual representation of the scheme for studying the influence of dispersion on temporal mode by measuring HOM type fourth-order interference of an unbalanced Mach-Zehnder interferometer. DL, delay line; BS, 50:50 beam splitter; D, detector; PZT, piezoelectric transducer; SMF, single-mode fiber.}
\label{Schematic_diagram}
\end{figure*}

Consider the scheme of a HOM interferometer shown in Fig.~\ref{Schematic_diagram}(a). For the sake of brevity,  suppose the fields are one-dimensional so we can absorb the position variable with time and only consider the temporal variable $t$. Assuming the two input fields of the beam splitter (BS), $E$ and $\bar{E}$, are initially in the same temporal mode (TM) but with the spectra of $\bar{E}$ altered by propagating through a medium with length corresponding to multiple pulse separations, the two input fields can be written as
\begin{align}
E(t)=\sum_{j} E_j(t-j\Delta T), ~~~\bar{E}(t)=\sum_{j} \bar{E}_j(t-(j+N)\Delta T),
\end{align}
where $\Delta T$ is the separation between adjacent pulses, and $N$ is the number of delayed pulses. Assume each pulse of the pulse train has an identical TM $\phi_k(t)$ but its amplitude and phase may fluctuate from pulse to pulse due to finite coherence time. Then $E_j(t)$ can be expressed as:
\begin{equation}
E_j(t) = A_je^{i\varphi_j} \phi_k(t),
\end{equation}
where
\begin{equation}
\phi_k(t)=\int^\infty_{-\infty}f_k(\omega)e^{-i\omega t}d\omega,
\label{Fourier}
\end{equation}
satisfying $\int dt |\phi_k(t)|^2=1$ is the $k$-th order TM with the duration of a single pulse $\delta t\ll\Delta T$. A set of temporal modes (TMs) are orthogonal with respect to a frequency (time) integral~\cite{Brecht-PRX2015}:
\begin{align}
\int^\infty_{-\infty}dt\phi_k^*(t)\phi_l(t)=\int^\infty_{-\infty}d\omega f_k^*(\omega) f_l(\omega)=\delta_{k,l}.
\end{align}
The field propagating through the delay $N\Delta T$ can be written as
\begin{align}
\bar{E}_j(t) = A_je^{i\varphi_j} \bar{\phi}_k(t)
\label{bar-E}
\end{align}
with
\begin{align}
\bar{\phi}_k(t)=\int^\infty_{-\infty}\bar{f}_k(\omega)e^{-i\omega t}d\omega, ~~~\bar{f}_k(\omega)=f_k(\omega)e^{-i\varphi(\omega) },
\label{bar-phi}
\end{align}
where $\varphi(\omega)$ denotes the phase shift induced by the delay of transmission medium.

In Fig.~\ref{Schematic_diagram}(a), the fields emerging at two outputs of BS are given by:
\begin{align}
E_1'(t)=[E(t)+\bar{E}(t+\Delta)]/\sqrt{2}, ~~~E_2'(t)=[\bar{E}(t+\Delta)-E(t)]/\sqrt{2},
\end{align}
where $\Delta$ is the additional adjustable delay between the fields $E$ and $\bar{E}$, which is obtained by passing one input field through a delay line (DL). The photocurrent of the detector (D1 and D2) can be expressed as:
\begin{align}
i_{D1}(t)=&\int_{T_R}d\tau k(t-\tau)E_{1}'^*(\tau)E_{1}'(\tau)\approx \frac{1}{2}\sum_j k(t-j\Delta T)I_j^{(1)},\nonumber\\
i_{D2}(t)=&\int_{T_R}d\tau k(t-\tau)E_{2}'^*(\tau)E_{2}'(\tau)\approx \frac{1}{2}\sum_j k(t-j\Delta T)I_j^{(2)},
\label{iD}
\end{align}
with
\begin{align}
I_j^{(1)}\equiv I_j+I_{j-N}+\Gamma_{j,N}(\Delta)+\Gamma_{j,N}^*(\Delta), ~~~I_j^{(2)}\equiv I_j+I_{j-N}-\Gamma_{j,N}(\Delta)-\Gamma_{j,N}^*(\Delta),
\label{I_j^{(1,2)}}
\end{align}
where $\Gamma_{j,N}(\Delta)\equiv\int d\tau E_j^*(\tau)\bar{E}_{j-N}(\tau+\Delta)$, and $k(t)$ is the response function of the detector with a response time of $T_R$. For the fields with pulse width $\delta t$ much smaller than $\Delta T$ and $T_R$, we have the approximation: $\int d\tau k(t-\tau)E_j^*(\tau-j\Delta T)E_i(\tau-i\Delta T)=0$ if $i\neq j$, and $\int d\tau k(t-\tau)\sum_j|E_j(\tau-j\Delta T)|^2\approx\sum_jk(t-j\Delta T)I_j$ with $I_j\equiv\int d\tau |E_j(\tau)|^2=|A_j|^2=\int d\tau |\bar{E}_j(\tau)|^2$. Moreover, we can add a piezoelectric transducer (PZT) in DL (not shown in Fig.~\ref{modes}(a)) so that the phase difference between two interfering fields fluctuates more than $2n\pi$ ($n$ is an integer much greater than 1) over a period $T$ covering many pulses. In this case, the second order interference terms $\Gamma_{j,N}(\Delta)$ and $\Gamma^*_{j,N}(\Delta)$ in Eq. (\ref{I_j^{(1,2)}}) are averaged to zero over the period $T$, i.e., $\langle\Gamma_{j,N}(\Delta)\rangle_T=0$, where $\langle\rangle_T$ is a pulse-to-pulse average. At each output of BS, the average intensity can be expressed as
\begin{align}
\langle i_{D1}\rangle_T=\langle i_{D2}\rangle_T=\frac{1}{2T}\int_T dt\sum_j k(t-j\Delta T)(I_j+I_{j-N})=R_pQ\langle I_j\rangle_T,
\end{align}
where $Q=\int dt k(t)$ is the total charge for one pulse, $R_p=1/\Delta T$ is the repetition rate of the pulsed field and $\langle I_j\rangle_T\equiv\frac{1}{M}\sum_j |A_j|^2$ is
the measured average intensity per pulse with $M\equiv T/\Delta T$ denoting the number of pulses to average. In general, the pulse-to-pulse average time $T$ is much longer than the delay time $N\Delta T$ so that $N\ll M$ and $\langle I_j\rangle_T \approx \langle I_{j-N}\rangle_T$.

Coincidence measurement between two detectors is
\begin{align}
R_c=\frac{1}{T}\int_T dt \int_{T_R} d(\Delta t)  i_{D1}(t)i_{D2}(t+\Delta t) =\frac{1}{4}R_p Q^2\langle I_j^{(1)}I_{j}^{(2)}\rangle_T,
\label{Rc}
\end{align}
here, we assume the detectors can resolve different pulses so that $T_R<\Delta T$ and $k(t-i\Delta T)k(t+\Delta t-j\Delta T)=0$ if $i\neq j$, and the delay between the two detectors $\Delta t$ is $0$.
Because of the fast scan of phase difference between $E$ and $\bar{E}$, the cross terms like $\langle I_j\Gamma_{j,N}\rangle_T $ and $\langle \Gamma^2_{j,N}\rangle_T$ etc. are averaged to zero, the fourth-order correlation term in Eq. (\ref{Rc}) can be written as
\begin{align}
\langle I_j^{(1)}I_{j}^{(2)}\rangle_T=\langle I_j^2\rangle_T+\langle I_{j-N}^2\rangle_T+2\langle I_j I_{j-N}\rangle_T-2\langle |\Gamma_{j,N}(\Delta)|^2 \rangle_T=2C\left[1-\xi\mathcal{V} (\Delta)\right],
\label{correlation}
\end{align}
where
\begin{align}
\mathcal{V} (\Delta)\equiv\left| \int dt \phi_k^*(t)\bar{\phi}_k(t+\Delta)\right|^2=\left| \int d\omega f_k^*(\omega)\bar{f}_k(\omega )e^{i\omega\Delta}\right|^2
\label{overlap}
\end{align}
describes the mode matching degree between the interfering fields $E$ and $\bar{E}$,
\begin{align}
\xi \equiv\frac{\langle|A_j|^2|A_{j-N}|^2\rangle_T}{\langle|A_j|^4\rangle_T+\langle|A_j|^2|A_{j-N}|^2\rangle_T}
\label{xi}
\end{align}
is determined by the intensities and photon statistics of two interference fields, and  $C\equiv\langle|A_j|^4\rangle_T+\langle|A_j|^2|A_{j-N}|^2\rangle_T$ is associated with the intensity fluctuation of the fields. Notice that this is a special case of a general scheme of unbalanced interferometers \cite{ou21}, the random phase is cancelled in quantity $|\Gamma_{j,N}(\Delta)|^2$ in Eq. (\ref{correlation}) and the result does not rely on the coherence of the input field.  

The visibility of fourth-order interference, i.e., the HOM dip, is
\begin{align}
V=\frac{R_c(\Delta\rightarrow\infty)-R_c(\Delta=0)}{R_c(\Delta\rightarrow\infty)}=\xi \mathcal{V}(0).
\label{visibility}
\end{align}
In Eq. (\ref{visibility}), we obviously have the overlapping integral $\mathcal{V} (0)=\int d\omega f_k^*(\omega)\bar{f}_k(\omega )=1$ if $f_k(\omega)=\bar{f}_k(\omega )$. In this condition, the visibility of fourth-order interference or the HOM dip $V$ is maximal, we have $V_{max}=\xi$. Eq. (\ref{xi}) shows that $\xi$ depends on the relative intensity and photon statistics of $E$ and $\bar{E}$. Since the intensities of $E$ and $\bar{E}$ are assumed to be equal, we have $\xi=1, 1/2, 1/3$ for the two fields in single photon state, coherent state and thermal state, respectively \cite{HOM1987,Ma2015}.

In general, it is difficult to obtain a thermal state or single photon state in a single TM~\cite{Grice1997,Ou1999,Brecht-PRX2015,su2019interference}. However, coherent state in single TM can be straightforwardly obtained by tailoring the output of a mode-locked laser with a wave shaper. To reveal how the transmission medium induced dispersion affects the spectra of TMs, hereinafter, we will focus on analyzing the model in Fig.~\ref{Schematic_diagram}(b), in which the two interfering fields of fourth-order interference are single mode coherent state.

In Fig. \ref{Schematic_diagram}(b), the fields $E$ and $\bar{E}$ are obtained by splitting the field in coherent state with a 50/50 beam splitter (BS1). So the scheme formed by two BSs in Fig. \ref{Schematic_diagram}(b) is an Mach-Zehnder interferometer (MZI). We study the spectra dependence of TM on dispersion when the difference of the fiber lengths in two arms $z$ corresponds to the delay of multiple pulses $N\Delta T$. The input field of MZI is in a single TM with order number $k$, the frequency components for the fields in two arms are related through the relation $\bar{f}_k(\omega)=f_k(\omega)e^{-i\beta(\omega)z }$, where $\beta(\omega) $ denotes the unbalanced dispersion induced by transmission fibers and can be described by the Taylor expansion
\begin{align}
\beta(\omega)=\beta_0+\beta_1(\omega-\omega_0)+\frac{1}{2}\beta_2(\omega-\omega_0)^2+\cdots.
\label{dispersion}
\end{align}
In Eq. (\ref{dispersion}), $\beta_m$ $(m=0,1,2...)$ denotes the $m$-th order dispersion coefficient. The zero order dispersion term is a constant
irrelevant to frequency. The first-order dispersion coefficient $\beta_1$
is related to the group-velocity through the relation $V_g=1/\beta_1$ and has no influence on the pulse shape, but the terms with $m\geq 2$ change the spectral or temporal feature of a pulsed field \cite{agrawal2000nonlinear}.

When the frequency bandwidth occupied by the TM is relatively narrow, the influence of higher order dispersion terms $\beta_m$ ($m\geq4$) on the evolution of pulsed field is negligibly small. In this case, the approximation $\beta(\omega)z\approx\frac{1}{2}(\omega-\omega_0)^2\beta_2 z= \varphi(\omega)$ holds, and the relation between $\bar{f}_k(\omega)$ and $f_k(\omega)$ can be rewritten as
\begin{align}
\bar{f}_k(\omega)=f_k(\omega)e^{-i\varphi(\omega) }.
\label{bar-fk(omega)}
\end{align}

To understand how the unbalanced dispersion in the MZI affects the mode matching of fourth-order interference, we simulate the mode overlapping factor $\mathcal{V} (0)$ (see Eq. (\ref{overlap})) when the TM of input field in Fig. \ref{Schematic_diagram}(b) takes different order number. In the simulation, a family of Hermite-Gaussian functions \cite{WalmsleyReview2020,Brecht-PRX2015}
\begin{align}
f_k(\omega)=i\frac{1}{\sqrt{2^kk!}}H_k\left(\frac{\omega-\omega_0}{\sqrt{2}\Delta\omega}\right)u(\omega)
\label{kth-gaussian}
\end{align}
with
\begin{align}
u(\omega)=\frac{1}{\sqrt{\Delta\omega}}\frac{1}{(2\pi)^{1/4}}exp\left[-\frac{(\omega-\omega_0)^2}{4\Delta\omega^2}\right]
\label{0th-gaussian}
\end{align}
denoting a Gaussian mean field mode, are successively substituted into Eqs. (\ref{bar-fk(omega)}) and (\ref{overlap}). In Eq. (\ref{0th-gaussian}), $\omega_0=\int\omega|u(\omega)|^2d\omega$ and $\Delta\omega^2=\int(\omega-\omega_0)^2|u(\omega)|^2d\omega$, are the mean value and variance of the field $u(\omega)$, respectively. It is straightforward to get the analytical expression of $\mathcal{V} (0)$ for TM with order number $k=0,1,2,3$:
\begin{align}
\mathcal{V} (0)=&\frac{1}{(1+B^2)^{1/2}}~(k=0), ~~~\mathcal{V} (0)=\frac{1}{(1+B^2)^{3/2}}~(k=1), \nonumber\\
\mathcal{V} (0)=&\frac{(2-B^2)^2}{4(1+B^2)^{5/2}}~(k=2), ~~~\mathcal{V} (0)=\frac{9(\frac{2}{3}-B^2)^2}{4(1+B^2)^{7/2}}~(k=3),
\label{overlap(0)}
\end{align}
where $B$ is related to the bandwidth of $0$-th order mode, second order dispersion $\beta_2$ and fiber length difference $z$ through the relation: 
\begin{align}
B=\Delta\omega^2\beta_2z.
\end{align}

\begin{figure*}[h]
\centering
\includegraphics[width=11cm]{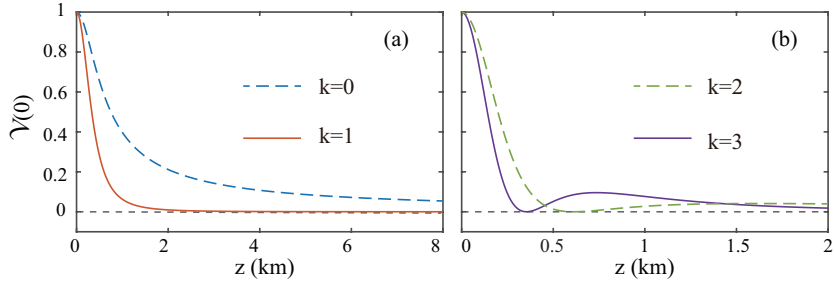}
\caption{The mode overlapping factor $\mathcal{V} (0)$ for the input temporal mode with order number of (a) $k=0,1$ and (b) $k=2,3$ when the fiber length difference in the two arms of Mach-Zehnder interferometer, $z$, is varied.}
\label{overlap-z}
\end{figure*}

The results in  Fig. \ref{overlap-z} are obtained by assuming the values of second order dispersion, the central wavelength, and bandwidth of $0$-th order mode are $\beta_2=-20$ $\mathrm{ps^2/km}$, $\lambda_0=2\pi c/\omega_0=1533$ nm and $\Delta\lambda_0=2\sqrt{2\ln2}\lambda_0\Delta\omega/\omega_0=1$ nm, respectively, where $c$ is the speed of light in vacuum.
Fig. \ref{overlap-z}(a) shows that for the case of $k=0,1$, $\mathcal{V} (0)$ always decreases with the increase of fiber length difference $z$, we have $\mathcal{V} (0)\rightarrow 0$ for $z\rightarrow \infty$. However, $\mathcal{V} (0)=0$, indicating the mode of $E$ and $\bar{E}$ is orthogonal to each other, can be achieved at some finite $z$ when $k\geq 2$. According to Eq. (\ref{overlap(0)}), for the case of $k=2,3$, we have $\mathcal{V} (0)=0$ under the condition of $|B|=\sqrt{2}$ and $|B|=\sqrt{2/3}$, respectively, as illustrated in Fig. \ref{overlap-z}(b).

\section{Experiments and Results}\label{Experiment and Results}

\begin{figure*}[h]
\centering
\includegraphics[width=\linewidth]{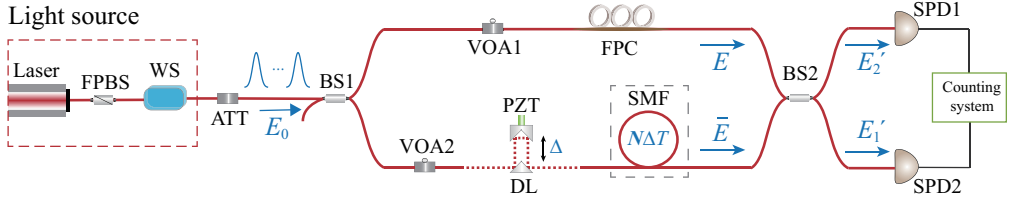}
\caption{Experimental setup. FPBS, fiber polarization beam splitter; WS, wave shaper; ATT, attenuator; BS1-BS2, 50:50 beam splitter; VOA1-VOA2, variable optical attenuator; PZT, piezoelectric transducer; DL, delay line; FPC, fiber polarization controller; SMF, single-mode fiber; SPD1-SPD2, single photon detector.}
\label{setup}
\end{figure*}

Our experimental setup is shown in Fig.~\ref{setup}. The input field of the unbalanced MZI $E_0$ having arbitrarily engineered TM profile is obtained by passing the output of a mode-locked fiber laser through a properly programmed wave shaper (WS, Finisar 4000A). The repetition rate and pulse duration of the laser are 50 MHz and 100 fs, respectively, and the central wavelength is in 1550 nm telecom band. To avoid the influence of self-phase modulation in transmission optical fiber on the spectral or temporal profile of the pulsed field  \cite{agrawal2000nonlinear}, the output of WS is heavily attenuated to the level of about one photon per pulse. After splitting the field $E_0$ into two by using BS1, the unbalanced dispersion in MZI is induced by propagating the two outputs of BS1 through standard single mode fibers (SMF) with length difference of $\sim 200$ m. Fiber polarization controller (FPC) placed in one arm is used to ensure the polarization of $E$ and $\bar{E}$ fields are well matched at BS2. The relative intensities of the fields in each arm is respectively adjusted by variable optical attenuator (VOA1 or VOA2), so that the intensities of two input fields of BS2 are equal. The two outputs of BS2 are respectively measured by the single photon detectors, SPD1 and SPD2.

The two SPDs (InGaAs-based) are operated in a gated Geiger mode. The 2.5-ns gate pulses coincide with the arrival of photons at SPDs. The response time of SPDs is about 1 ns, which is  100 times longer than the pulse duration of the detected field. The electrical signals produced by the SPDs in response to the incoming photons are reshaped and acquired by a computer-controlled analog-to-digital (A/D) board. So the individual counting rate of two SPDs, $R_1$ and $R_2$, and two-fold coincidences acquired from different time bins can be determined because the A/D card records all counting events. The fourth-order interference of the fields $E$ and $\bar{E}$ is measured by the coincidence rate originated from the same time bin, $R_c$. During the measurement, the PZT mounted on DL is scanned at a rate of 30 Hz. Under this condition, the second-order coherence effect between $E$ and $\bar{E}$ fields is averaged out. The counting rates of $R_1$ and $R_2$ recorded in the period $T$ of one second stay constant, while the two-fold coincidence counting rate $R_c$ varies with $\Delta$. For clarity, the results of $R_c$ is normalized by the accidental coincidence rate $R_{acc}$, which is the coincidence originated from adjacent time bins and is equivalent to $R_c(\Delta\rightarrow\infty)$. According to Eq. (\ref{correlation}), the normalized coincidence is related to the mode overlapping factor $\mathcal{V} (\Delta)$ through the realtion:
\begin{align}
    N_c(\Delta)=\frac{R_c}{R_{acc}}=\frac{R_c/R_p}{(R_1/R_p)(R_2/R_p)}=1-\xi\mathcal{V} (\Delta),
    \label{Nc}
\end{align}
where $\xi=1/2$ because the input of MZI is in coherent state and the ideal mode match between $\bar{E}$ and $E$ is achievable.

\begin{figure*}[h]
\centering
\includegraphics[width=12cm]{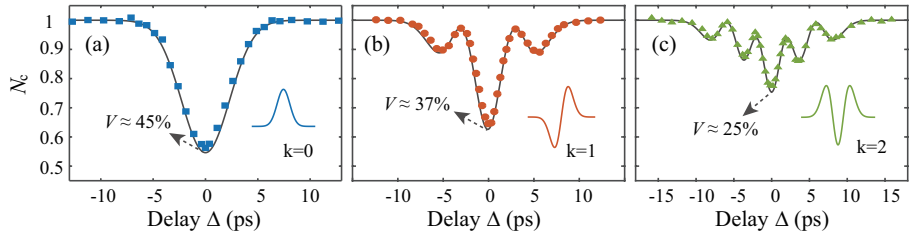}
\caption{The normalized coincidence $N_c$ as a function of $\Delta$  when the input field of the unbalanced MZI is in the single temporal mode with (a)$k=0$, (b)$k=1$, and (c)$k=2$, respectively. The solid curves in each plot are the theoretical simulate results.}
\label{orth_g2}
\end{figure*}

We first measure the coincidence rate $R_c$ by varying the delay $\Delta$ when the input field $E_0$ of the unbalanced MZI is respectively in a set of TM with different order number $k$. In the experiment, the central wavelength of $E_0$ is  fixed at $\lambda_0=1533$ nm, and the full width at half maximum (FWHM) of the intensity spectrum $|f(\omega)|^2$ of 0-th order Gaussian shaped beam is $\Delta\lambda_0=1$ nm. Figs. \ref{orth_g2}(a), \ref{orth_g2}(b) and \ref{orth_g2}(c) plot the data of normalized coincidence $N_c=R_c/R_{acc}$, which are obtained when the order number of TM is $k=0,1,2$, respectively. The solid curves in Fig. \ref{orth_g2} are obtained by substituting the experimental parameters into Eqs. (\ref{Nc}), (\ref{overlap}), (\ref{xi}), and (\ref{bar-fk(omega)})-(\ref{0th-gaussian}), showing the theory predictions agree well with the experimental data. We find $N_c(\Delta)$ in each case is symmetric around the zero-delay point $\Delta=0$, but the pattern depend on the mode order number. $N_c(\Delta)$ monotonously increases with $|\Delta|$ for $k=0$. However, there exists oscillation in the pattern of $N_c(\Delta)$ for higher order mode, and the number of the oscillation peaks in each side is the same as order number $k$ due to the multiple peak structure of $\phi_k(t)$. The visibility of HOM dip is $V=45\%, 37\%, 25\%$ for the case of $k=0,1,2$ respectively. Obviously, the visibility of HOM is always less than the ideal value $50\%$, indicating the spectral or TM matching between $E$ and $\bar{E}$ is altered by the unbalanced dispersion in the two arms of MZI. Moreover, one sees that the visibility $V$ decreases with the increase of order number $k$, which means that the degree of mode mismatch increases with $k$. This is because that for a set of TM, which forms a field-orthogonal basis, the frequency bandwidth occupied by the TM increases with the order number, as illustrated in Fig. \ref{modes}(c).  As a result, for a fixed fiber length difference $z$, the amount of unbalanced dispersion increases with $k$.

\begin{figure*}[h]
\centering
\includegraphics[width=12cm]{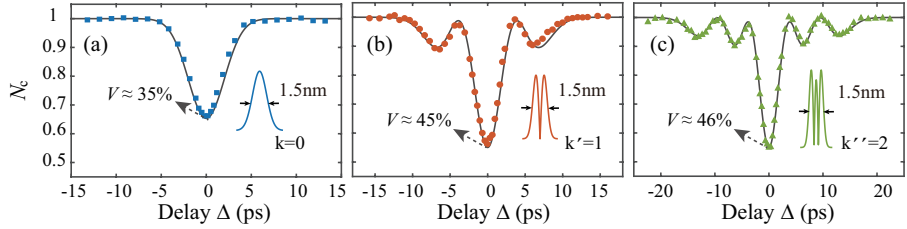}
\caption{The normalized coincidence $N_c$ as a function of $\Delta$ when the input temporal mode of the unbalanced MZI occupies the same frequency resources but  the order of temporal mode is (a)$k=0$, (b)$k^{\prime}=1$, and (c)$k^{\prime\prime}=2$, respectively. The solid curves in each plot are the theoretical simulate results.}
\label{eqBW_g2}
\end{figure*}

We then repeat the measurement of $N_c(\Delta)$ when the frequency band occupied by input field $E_0$ is the same. In the experiment, the central wavelength of $E_0$ is still $\lambda_0=1533$ nm, but the FWHM of $E_0$ field, having the order number $k=0$, $k^{\prime}=1$ and $k^{\prime\prime}=2$ respectively belonged to different mode basis, is fixed at 1.5 nm. The data for the cases of $k=0$, $k^{\prime}=1$ and $k^{\prime\prime}=2$ is shown in Figs.~\ref{eqBW_g2}(a), \ref{eqBW_g2}(b) and \ref{eqBW_g2}(c), respectively. Also, we simulate the results by substituting the experimental parameters into Eqs. (\ref{Nc}), (\ref{overlap}), (\ref{xi}), and (\ref{bar-fk(omega)})-(\ref{0th-gaussian}), as shown by the solid curves in Fig.~\ref{eqBW_g2}, which well agree with the data points. We find the visibility of HOM dip is $35\%$, $45\%$ and $46\%$ for $k=0$,$k^{\prime}=1$ and $k^{\prime\prime}=2$, respectively. Different from Fig.~\ref{orth_g2}, here, the visibility $V$ increase with the order number.

\begin{figure*}
\centering
\includegraphics[width=7cm]{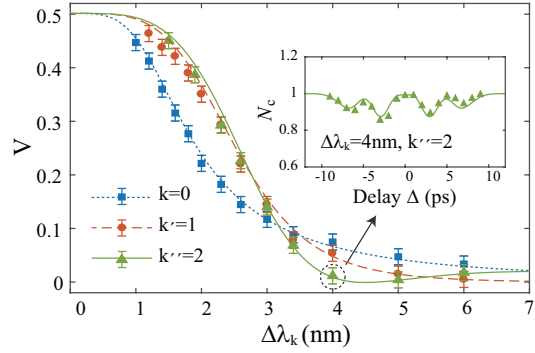}
\caption{The visibility of HOM dip $V$ as the functions of the FWHM of $k$-th mode $\Delta\lambda_k$, and the order of TM $k$. The inset plots the normalized coincidence $N_c$ versus delay $\Delta$ when the bandwidth and order number of TM are $\Delta\lambda_k=4$ nm and $k^{\prime\prime}=2$, respectively.}
\label{visibility_bandwidth}
\end{figure*}

To further understand how the dispersion induced by transmission medium affect the spectral profile of TM, we measure $N_c$ as a function of $\Delta$ and accordingly deduce the visibility $V$ of fourth-order interference when the bandwidth of input field is changed. In this experiment, the central wavelength of input $E_0$ is still fixed at $\lambda_0=1533$ nm, but the FWHM of $E_0$ is changed from 1.5 nm to 6 nm. For each setting of FWHM, the spectra profile of $E_0$ is tailored into $k=0$, $k^{\prime}=1$, or $k^{\prime\prime}=2$ of different TM basis. As shown in Fig.~\ref{visibility_bandwidth}, the blue squares, orange circles, and green triangles correspond to the data of $V$ for $k=0$, $k^{\prime}=1$ and $k^{\prime\prime}=2$, respectively. We also simulate the visibility $V$ by substituting the experimental parameters into Eqs.~(\ref{visibility}) and (\ref{overlap(0)}), as shown by the dotted, dashed and solid curves in Fig.~\ref{visibility_bandwidth}. We find that when the bandwidth of input is relatively narrow, the visibility for $k=0$ is the lowest. With the increase of the bandwidth, the visibility for $k=0$ starts to become higher than that for the modes occupying the same frequency band but with higher order mode number. Moreover, for the case of $k=0$ and $k^{\prime}=1$, $V$ decreases with the increase of bandwidth. While for the case of $k^{\prime\prime}=2$, we have $V\approx0$ for $\Delta\lambda_{k^{\prime\prime}=2}=4$ nm. The inset in Fig.~\ref{visibility_bandwidth} plots $N_c$ versus $\Delta$ for $\Delta\lambda_{k^{\prime\prime}=2}=4$ nm. One sees that although the patten of $N_c(\Delta)$ shows the visibility $V$ is approaching $0$, there exists oscillation structure symmetrically distributed around the central point $\Delta=0$. The result indicates that when the FWHM of $E_0$ field is 4 nm, $E$ and $\bar{E}$ with $k^{\prime\prime}=2$ are about orthogonal at $\Delta=0$, but the overlap between the two interfering fields may increase when $\Delta$ is away from $\Delta=0$ due to the multiple peak structure of higher order TM.

\begin{figure*}[h]
\centering
\includegraphics[width=7cm]{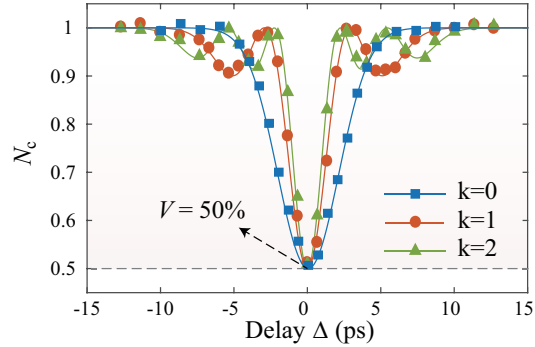}
\caption{The normalized coincidence $N_c$ measured by varying the TM $\phi_k (k=0,1,2)$ and the relative delay $\Delta$ between $E$ and $\bar{E}$ with a balanced MZI. The solid curves in each plot are the theoretical simulate results.}
\label{Balance_g2}
\end{figure*}

Finally, we modify the unbalanced MZI into a balanced one and measure $N_c$ as a function of $\Delta$ when the input field $E_0$ is the same as in observing Figs.~\ref{orth_g2}. In this experiment, the balanced MZI with $z=0$ is realized by either inserting a 200 m long standard SMF in the arm of $E$ field or taking out the 200 m SMF in the arm of $\bar{E}$ field. In this case, we have $f_k(\omega)=\bar{f}_k(\omega)$, which means that the mode profiles of $E$ and $\bar{E}$ are identical. In Fig.~\ref{Balance_g2}, the data represented by the blue squares, orange circles, and green triangles are the measured $N_c(\Delta)$ for a set of TM with $k=0,1,2$, respectively. Again, the solid curves are simulation results obtained by substituting experimental parameters into Eqs. (\ref{Nc}), (\ref{overlap}), (\ref{xi}), and (\ref{bar-fk(omega)})-(\ref{0th-gaussian}). As expected, the simulations agree well with experimental results. In each case, the visibility of observed HOM dip is $V=50\%$. The results imply that for homodyne detection (HD) used in the information processing~\cite{huo-PRL-20}, the propagation induced distortion of the spectral or temporal profile of probe field can be compensated by passing the local oscillator of HD through the same propagation medium. Otherwise, the distortion will cause a reduced detection efficiency owning to the mode mismatching.

\section{Discussion}\label{Discussion}

In Sec.~\ref{Theory} and Sec.~\ref{Experiment and Results}, the agreement between the experimental data and theoretical expectation indicates the dispersion parameter of transmission fiber $\beta_m$ can be obtained by fitting the data of measured $N_c(\Delta)$. Indeed, it is well known
that the pattern of the HOM type fourth-order interference depends on the indistinguishability of two interfering fields~\cite{HOM1987,Ou1999}. The broadening of HOM dip in the pattern of $N_c(\Delta)$ and the decrease of the visibility due to unbalanced dispersion have already been analyzed and observed when the two Gaussian shaped fields are in thermal state~\cite{Ma2011}, coherent state~\cite{Zhouqiang2021-PR} and single photon state~\cite{Kim2021,Ma2015}, respectively. Moreover, it has been shown that the second-order
dispersion coefficient of transmission fiber can be measured by characterizing the broadening effect of the fourth-order interference pattern~\cite{Zhouqiang2021-PR}. However, the investigations previously reported focus on the Gaussian shaped fields, the relation between the fourth-order interference pattern and the interfering fields in higher order temporal modes has not been studied.

From Eqs. (\ref{Nc}), (\ref{bar-fk(omega)})-(\ref{overlap(0)}), we find the normalized coincidence $N_c(\Delta)$ for the TM with $k=0$ is related with $B=\Delta\omega^2\beta_2 z$ through the formula:
\begin{align}
N_c(\Delta)=1-\frac{\xi}{(1+B^2)^{1/2}}\exp{(\frac{-\Delta\omega^2\Delta^2}{1+B^2})}=1-V\exp{(\frac{-\Delta\omega^2\Delta^2}{1+B^2})}.
\end{align}
According to the definition of the FWHM ($\Delta\tau$) of HOM dip $N_c(\pm\frac{\Delta\tau}{2})=\frac{N_c{(\Delta\rightarrow\infty)}+N_c{(\Delta=0)}}{2}=1-\frac{V}{2}$,
it is straightforward to deduce the relation between $\Delta\tau$ and the visibility of fourth-order interference V:
\begin{align}
\Delta\tau=\frac{2\xi\sqrt{\ln 2}}{\Delta\omega V}.
\end{align}
Therefore, in the process of measuring dispersion coefficient of transmission line in one arm of MZI, instead of characterizing the full pattern of $N_c(\Delta)$ to extract out its FWHM, we can simply evaluate the visibility by measuring $N_c(\Delta=0)$~(or $R_c(\Delta=0)$). Obviously, the procedure of measuring $N_c(\Delta=0)$ is much fast and simple. 

When the bandwidth of input field is narrow, the influence of $\beta_m$ with $m\geq 3$ on the interference pattern is negligible, we can deduce coefficient $\beta_2$ in one arm of unbalanced MZI from the visibility of fourth-order interference. The effect of $\beta_m$ with $m\geq 3$ will come in when we enlarge the bandwidth of input field and we can measure them by increasing the bandwidth. However, the method is somewhat complicated and we will discuss the detail elsewhere. Once the different order dispersion coefficients $\beta_m$ ($m\geq 2$) of the transmission line with the channel centered at a given wavelength are available, the spectral distortion of the TM can be recovered by compensating the dispersion with a wave shaper ~\cite{pulikkaseril2011spectral,schroder2013optical}.

Comparing with the methods which are based on the second-order interference and often used to measure the dispersion of transmission medium, such as the white-light interferometer method and the modulation phase shift method etc.~\cite{lee2006versatile,cohen1985comparison}, this method has the following features: (1) the value of all higher order dispersion coefficient $\beta_m$ ($m\geq 2$) at the central wavelength of $E_0$ field can be directly obtained; (2) the length of the transmission fiber to be measured can be in the range of a few meters to at least tens kilometres.

As seen from Eq. (\ref{overlap}), the visibility only depends the overlap of the temporal mode functions of two interfering fields but not on the phase correlation between them. Thus, it is not sensitive to coherence time of the original input field to the MZI. Indeed, HOM interference was observed between two totally independent pulsed fields with no phase relation at all \cite{Ma2011,su2019interference}. So, this technique is not limited to the length of delay. Indeed, the unbalanced MZI scheme used here is a special case of a more general scheme of unbalanced interferometers~\cite{ou21} where the unbalanced optical path is much larger than the coherence length of the interfering field and yet we can still observe interference in coincidence measurement.

\section{Conclusion}\label{Conclusion}

In conclusion, in order to demonstrate how the dispersion induced by transmission optical fiber influence the spectral profile of different order TMs, we study the fourth-order interference of an unbalanced MZI. In the MZI, the two interfering fields of HOM interferometer are originally from one single temporal mode field but respectively propagate through two pieces of optical fiber to achieve unbalanced dispersion. The amount of unbalanced dispersion can be changed by changing the fibre length in the two arms of MZI. Both the simulation and experimental results show that the interference pattern of HOM type fourth-order interference varies with the order number of TM. For a set of TM described by Hermite-Gaussian function, the interference pattern of the $0$-th order TM is Gaussian shaped; while for the higher order mode, there exist oscillation in each side of the zero-delay point, and the number of peaks in the oscillation of one side is the same of order number $k$. Moreover, we find the visibility of fourth-order interference reflects the mode mismatching induced by unbalanced dispersion, and the relation between the mode mismatching degree and the amount of unbalanced dispersion varies with the order number and frequency bandwidth of TM. The visibility monotonically decrease with the amount of unbalanced dispersion when the order number of TM is $k=0,1$. However, for the case of $k\geq 2$, the visibility can be equal to zero when the unbalanced dispersion is modest, which means that the two interfering fields become orthogonal. In addition, we find the fourth-order interference of unbalanced MZI with the input of a single temporal mode coherent state can be used to measure the dispersion of transmission optical fiber. In particular, by simply evaluating the deviation of visibility from the maximal value of $50 \%$, the dispersion coefficients can be directly measured. 
Our investigation is useful for further studying the distribution of temporally multiplexed quantum states in fiber network~\cite{Brecht-PRX2015,huo-PRL-20}.

\begin{backmatter}
\bmsection{Funding}
This work was supported in part by National Natural Science Foundation of China (Grants No. 91836302, No. 12074283, and No. 11874279) and Science and Technology Program of Tianjin (Grant No. 18ZXZNGX00210).


\end{backmatter}



\end{document}